\newcommand{\BZ}{\mathrm{BZ}}
\newcommand{\dk}{\,d\mathbf{k}}
\newcommand{\Ekn}{E_{n}(\mathbf{k})}
\newcommand{\projamp}[2]{\bigl|\langle #1 \mid #2 \rangle\bigr|^2}

\UseRawInputEncoding
\documentclass[sn-mathphys]{sn-jnl}
\usepackage{amsmath,amssymb,amsfonts,bm}
\usepackage{mathrsfs}

\usepackage{graphicx}
\usepackage{multirow}
\usepackage{booktabs}
\usepackage{algorithm}
\usepackage{algorithmicx}
\usepackage{algpseudocode}
\usepackage{listings}

\usepackage[dvipsnames]{xcolor}
\usepackage[textsize=tiny]{todonotes}

\theoremstyle{thmstyleone}%

\theoremstyle{thmstyletwo}%

\theoremstyle{thmstylethree}%

\begin{document}

\title[Quasi-van Hove informed DOS in GNN]{Quasi-van Hove singularities informed approach improving DOS/pDOS predictions in GNN}

\author*[1]{\fnm{Grigory} \sur{Koroteev}}\email{grigory.koroteev@mbzuai.ac.ae}\email{greg.koroteev@gmail.com}
\affil*[1]{\orgdiv{MBZUAI}, \orgaddress{\city{Abu Dhabi}, \country{UAE}}}

\abstract{%
We propose a quasi–van Hove–informed refinement for density-of-states (DOS/pDOS) prediction in graph neural networks (GNNs). The method augments a baseline model with a peak-aware additive component whose amplitudes and widths are optimized under a cosine–Fourier loss with curvature and Hessian priors, yielding improved DOS fidelity in both cosine distance and MAE metrics.
}


\maketitle








\section{Introduction}\label{sec1}

The endeavor to model quantum systems using machine learning techniques is both challenging and captivating. A related but separate question is whether a precise mathematical description of such systems is achievable at all.  In recent years, there has been a surge in research related to Physics-Informed Neural Networks \cite{Karniadakis2021PIML} \cite{Wang2022PINNsFail}, particularly in the area of complex loss functions that emulate real physical equations. 

In this paper, we propose an enhancement to the GNN model for predicting single-particle perturbations. The approach to post-processing is presented, which significantly improves the final result of m-layer (here $m=3$) machine learning by incorporating information evaluated at the first layer. This enhancement is based solely on the general properties of numerical solutions and general physical reasoning. The efficacy of the proposed method is demonstrated through its application to diverse quantum systems, from molecules to glasses, using Materials Project data \cite{Jain2013MaterialsProject},\cite{Horton2025MPPerspective}. Symbolically: Dataset = MP [3, 4], $N_{samples} ≈ 32k$.

\section{Subject of investigation}

In the most general terms, we need to define the subject of modeling from both a physical and a machine learning perspective. The density of states (DOS) spectrum is typically defined as the number of quantum states per unit energy interval. It represents the distribution of available states as a function of energy, which can be thought of as the energy levels that would be occupied or released in transitions involving the breaking of quantum interactions or perturbations in the system. The dimensionality of this quantity is given by $[states/eV]$. A detailed description requires a specific physical model: at least specifying the type of particles (or quasi-particles) through which the interaction within the system takes place, the degree of degeneracy, and so on. Let us first recall how one-particle states are represented:


\begin{equation}\label{D(E,psi)}
    D(\vec{r}, E) = \sum_n |\psi_n(\vec{r})|^2 \, \delta(E - E_n)
\end{equation}

in general or electronic case, where $\psi_n(\vec{r})$ is a space-dependent wavefunction of the $n-$th state, $E_n$ is energy $n-$excitation $D_{\text{total}}(E) = \sum_\alpha D_\alpha(E)$ \cite{Kittel2005ISSPhys}.   


To describe crystalline systems, it is often more convenient to work with densities per unit volume to allow for direct integration over the Brillouin zones, where wavevectors $\vec{k}$ are used \cite{vanHove1953Singularities}.  

$$D(E) = \sum_n \int_{\text{BZ}} \delta(E - E_n(\vec{k})) \, \frac{d\vec{k}}{(2\pi)^3} $$

An exact description of quantum systems is still a dream, but some fundamental points for machine learning tasks can be noted. For example, the more complex the system, the more collective effects need to be considered, which will be finite continuous functions in terms of the probability density distribution. At the same time, individual states will still be locally independent and can be described similarly to delta functions. However, neither the real resolving power of devices, nor numerical methods, nor the multitude of subtle physical effects allow working with delta functions. In fact, it is a matter of working with high peaks. It should also be remembered that in practice one or another type of smoothing is usually applied. Graphically, this means that, to a first approximation, we are dealing with a combination of continuous functions and individual peaks.

\section{Peaks candidates}


A common consequence of numerical methods, especially machine learning, is the smoothing of the resulting functions. This is notably critical for delta functions, which are much more sensitive to random errors and smoothing effects. Knowing this, there is an understandable desire to capture some of the information that may indicate peaks.

If we were discussing well-studied field of knowledge, such as solid state physics \cite{vanHove1953Singularities},\cite{AshcroftMermin1976} or superconductors \cite{Lynn1990HTSC}, we would be able to present precise analytical model-dependent calculations. As an illustration of the idea, we will use a classic example of how DOS is described in crystals.

\begin{equation}
    D(E) = \frac{(2m_{DOS}^*)^{3/2}}{2\pi^2 \hbar^3} \sqrt{|E-E_b|}
\end{equation}

where $m^*$ is an effective mass, $ E_b $ is band edge energy  \cite{Kittel2005ISSPhys}.  According to theory, 

VHS at $ \nabla_k E = 0 $, leading to $ D(E) \propto \log|E - E_c| $ \cite{vanHove1953Singularities}. However, a thorough analysis of the nuances and applicability of the model is beyond the scope of this article.

Therefore, we will consider quasi-Van Hove singularities (qVHS) of the $ D_{total}(E-E_{Fermi}) = \Sigma_j D_j(E-E_{Fermi})    $ function as  candidates to find the peaks of the DOS distribution. 

Let us determine the $ \mathcal{GNN}k[F(x)] $ as GNN representation of the $ F(x) $ function after the $k-$th layer. 

$\mathbf{ Statement}\label{Statement}$ : 

Candidates for qVHS can be estimated as the zeros of the following equations and can play a role of "lost" peaks:
\begin{equation}\label{eq:second}
    \frac{\partial }{\partial E}{\mathcal{GNN}1[D_{total}}(E-E_{Fermi})] = 0 , 
\end{equation}

In other terms, it makes sense to look for candidates with the following properties among critical points:: 

\begin{equation}
    \mathcal{GNN}1[D_{total}(E-E_{Fermi})]  \longrightarrow  \frac{\partial }{\partial E}{\mathcal{GNN}1[D_{total}}(E-E_{Fermi})] = 0  
\end{equation}

$$
      PeaksCandidates\bigl\{ E_0,E_1, ... , \bigl\}
$$ 


The method adds unique, physically motivated information per sample via BHL optimization.
Next, we will explain why using the Bayesian approach to determine the peak parameters is important. To make the picture clearer, we will now describe the computational experiment and show its results.

\begin{figure}
    \centering
    \includegraphics[width=0.5\linewidth]{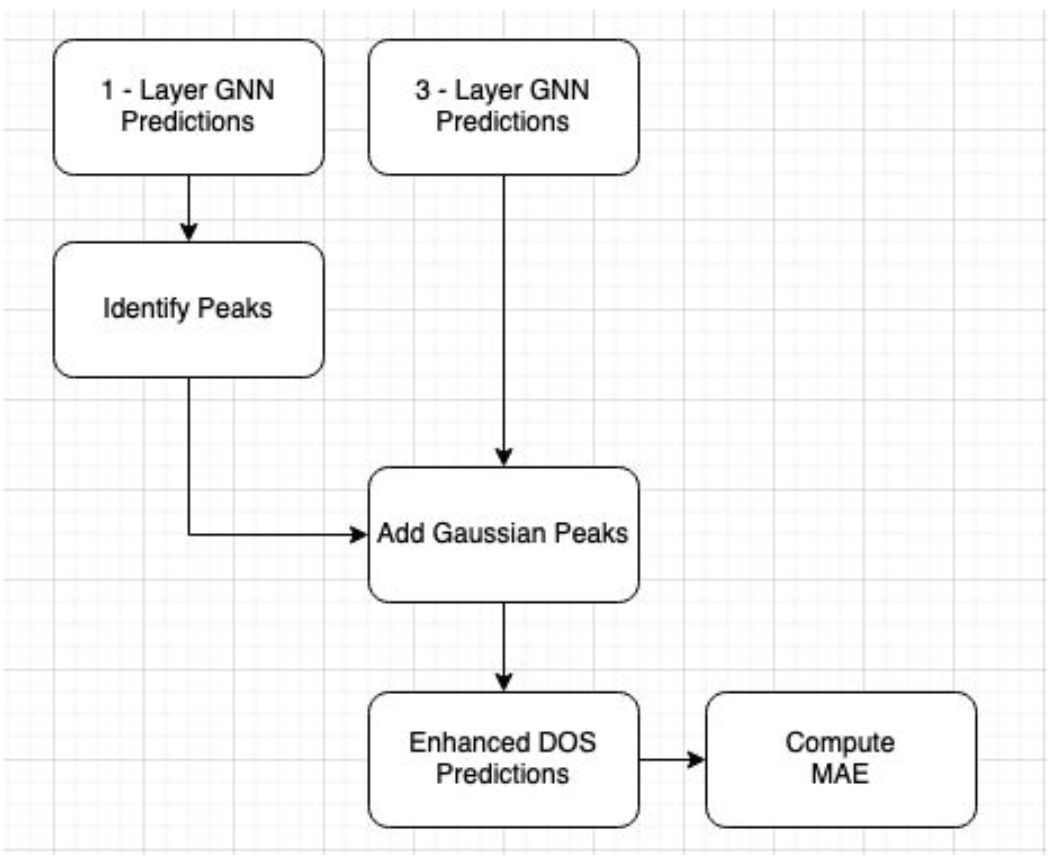}
    \caption{The principal scheme of proposed process. After 1-layer GNN we started find quasi-VHS to use them at the final stage.} 
    \label{fig:Scheme}
\end{figure}

First, a detailed description will be provided of the methodology to be employed for the application of the peaks that have been identified, as well as a rationale for why these peaks contain sufficient physical information to be applicable in machine learning.



\subsection{Projected DOS (pDOS)}

It is important to note that this method is not only applicable to the full spectrum, but also to its projections, such as those projected onto the atom. The so-called pDOS can be regarded as conditionally independent and subsequently amalgamated into a comprehensive distribution of state densities.

Let $\{\phi_{a\mu}\}$ denote localized basis functions (orbital channel $\mu$ on atom $a$).
For Bloch eigenstates $\psi_{n\mathbf{k}}$ with energies $\Ekn$, define the per-orbital pDOS
\begin{equation}
  D_{a\mu}(E)
  \;=\;
  \sum_{n}\int_{\BZ}
  \projamp{\phi_{a\mu}}{\psi_{n\mathbf{k}}}\;
  \delta\!\bigl(E-\Ekn\bigr)\;
  \frac{\dk}{(2\pi)^d}.
\end{equation}
The \emph{atomic} pDOS is the sum over channels on atom $a$,
\begin{equation}\label{eq7}
  D_{a}(E)\;=\;\sum_{\mu\in\Lambda_a} D_{a\mu}(E),
\end{equation}
and the total DOS decomposes as
\begin{equation}\label{eq8}
  D_{\mathrm{tot}}(E)\;=\;\sum_{a} D_{a}(E)
  \;=\;\sum_{a}\sum_{\mu\in\Lambda_a} D_{a\mu}(E).
\end{equation}
(If spin is resolved, add an index $s$ and sum over $s$.)

In what follows, we employ a notation for denoting the density of states or its projections that features a minimal set of indices, subject to the following conditions \ref{eq7} or \ref{eq8}. Nonlinear behavior can be generalized, but this is beyond the scope of this article.

\section{quasi van Hove Singularities (qVHS)}

A distinguishing feature of the task under consideration, the application of supplementary physical information to enhance the final estimates of physical quantities (particularly the DOS/pDOS), is that it is, in essence, antithetical to the primary task of standard DFT. In the domain of quantum chemistry, a model is constructed to illustrate the interaction of atoms or electrons. Subsequent to determining the dispersion relations, these relations are converted into a spectrum of states. In our task, we have already obtained an estimate of this distribution of DOS based on the results of machine learning, despite our lack of knowledge regarding the dispersion relations. This ability to manipulate predictions enables the refinement of learning methods.  In order to facilitate a meaningful comparison, it is necessary to have at least two relatively independent predictions.

In this study, we propose an approach to training the GNN model in one metric in two distinct variants independently. The number of layers is equivalent to the maximum attainable prediction quality, excluding post-processing. To illustrate this point, we will employ the PaiNN \cite{Schutt2021PaiNN} model and assume that a 3-layer scheme is optimal for it.  We will discuss the choice of metrics and normalizations in the next section.

Let us return to the logic in which the expression for VHS \cite{vanHove1953Singularities} was originally obtained. DOS could be presented as $D(E)$:


\[
D(E)
\;=\;
\int_{S(E)} \frac{dS}{\bigl\lvert\nabla_{\bm{k}}E(\bm{k})\bigr\rvert},  \mathcal{D}[E](\mathbf{k}) = \frac{1}{|\nabla_{\mathbf{k}} E(\mathbf{k})|}
\]
where \(S(E)=\{\bm{k}:E(\bm{k})=E\}\) is the constant‐energy surface.

All the available information we can use here is estimates of the spectrum of states and its derivatives. The critical points of the equation will be presented in three different forms: equation \ref{eval_9} as integral form, \ref{eval_10}: divergence expansion, \ref{eval_11}: $\mathcal{D}'[E]$ operator. This will allow for analysis of the behavior of the functions at these points from several sides.
Additionally, sufficient conditions will be indicated for choosing such points as candidates for qVHS:

\begin{equation}\label{eval_9} 
    \frac{d D(E)}{dE} = \int_{E(\mathbf{k}) = E} \nabla_{\mathbf{k}} \cdot \left( \frac{\nabla_{\mathbf{k}} E}{|\nabla_{\mathbf{k}} E|^2} \right) \, dS  = 0 
\end{equation}

Zeros of integrand expression :

\begin{equation}\label{eval_10}
    \nabla_{\mathbf{k}} \cdot \left( \frac{\nabla_{\mathbf{k}} E}{|\nabla_{\mathbf{k}} E|^2} \right)
=
\frac{\nabla^2_{\mathbf{k}} E}{|\nabla_{\mathbf{k}} E|^2}
- 2 \frac{ (\nabla_{\mathbf{k}} E)^\top \cdot \nabla^2_{\mathbf{k}} E \cdot \nabla_{\mathbf{k}} E }{|\nabla_{\mathbf{k}} E|^4} = 0
\end{equation}



In the form of a nested operator \(\mathcal{D}'[E](\mathbf{k}) = -\nabla_{\mathbf{k}} \cdot \left( \frac{\nabla_{\mathbf{k}} E}{|\nabla_{\mathbf{k}} E|^2} \right) \cdot \frac{1}{|\nabla_{\mathbf{k}} E|}\) such that:

\begin{equation}\label{eval_11}
\frac{dD}{dE} = \int_{E(\mathbf{k}) = E} \mathcal{D}'[E](\mathbf{k}) \, dS
\end{equation}

\subsection{Critical Points and \(\nabla_{\mathbf{k}}E=0\)}

Points where
\[
\nabla_{\mathbf{k}} E(\mathbf{k})=0
\]
(called critical or stationary points in the energy dispersion) are essential for the DOS. Near such points, the energy dispersion can be approximated via a second-order Taylor expansion around the critical point \(\mathbf{k}_0\):
\[
E(\mathbf{k}) \approx E(\mathbf{k}_0) + \frac{1}{2}(\mathbf{k}-\mathbf{k}_0)^T \mathbf{H}_E(\mathbf{k}_0)(\mathbf{k}-\mathbf{k}_0),
\]
where the Hessian matrix is defined as:
\[
[\mathbf{H}_E(\mathbf{k}_0)]_{\mu\nu} = \frac{\partial^2 E(\mathbf{k}_0)}{\partial k_{\mu}\partial k_{\nu}}.
\]

At this stage, the possibility of checking the distribution of the signs of the eigenvalues over the components of the vector $\textbf{k}$ is not accessible. However, this turns out to be unnecessary, especially considering that the approach used to describe systems with very different, and generally unknown, dispersion equations is employed.

We consider the condition
\[
\nabla_{\mathbf{k}} \cdot \left( \frac{\nabla_{\mathbf{k}} E}{|\nabla_{\mathbf{k}} E|^2} \right) = 0,
\]
which determines special loci in $\mathbf{k}$--space where the curvature balance of the band dispersion $E(\mathbf{k})$ enforces qualitative changes in the density of states (DOS). The interpretation can be given in the framework of catastrophe theory and Morse theory \cite{arnold1985singularities},\cite{milnor1963morse}:

\begin{itemize}
  \item \textbf{Stationary DOS structures:}  
  Zeros of the above expression correspond to stationary points of $\tfrac{dD(E)}{dE}$, indicating where the topology of isoenergy surfaces undergoes change.

  \item \textbf{Catastrophe-theoretic viewpoint:}  
  In catastrophe theory, singularities arise when the Jacobian of the mapping $\mathbf{k} \mapsto E(\mathbf{k})$ degenerates.  
  Here, $\nabla_{\mathbf{k}} E = 0$ identifies van Hove singularities.  
  The further balance condition $\nabla_{\mathbf{k}} \cdot \bigl(\nabla_{\mathbf{k}} E / |\nabla_{\mathbf{k}} E|^2 \bigr) = 0$ selects higher-order bifurcation points.

  \item \textbf{Arnold’s classification:}  
  According to Thom--Arnold classification \cite{thom1975structural}, such singularities correspond to canonical catastrophes:  
  \begin{itemize}
    \item \emph{Fold (A$_2$):} simple extremum of $E(\mathbf{k})$,
    \item \emph{Cusp (A$_3$):} merging of two folds, realized in higher-order van Hove points,
    \item \emph{Swallowtail and umbilics (D-series):} appear when multiple Hessian eigenvalues vanish simultaneously.
  \end{itemize}

  \item \textbf{Physical meaning:}  
  These zeros mark energy values where the topology of the Fermi surface changes \cite{lifshitz1960anomalies}, \cite{blanter1994electronic} (Lifshitz transitions) and where DOS develops non-analytic behavior.
\end{itemize}

Consequently, a proposal is hereby made for the utilization of these features as qVHS. In the most fundamental sense, the information for each point can be characterized as a Gaussian peak:

\begin{equation}\label{hessian}
    H_i = -\frac{A_i}{\sigma_i^2} \delta_i^{\lambda}
\end{equation}

where $A_i$ is the amplitude of the found peak, $\sigma_i$  - the peak width and $ \delta_i^{\lambda} = 1$ if $i$ -th peak with energy $E_i$ cannot be excluded by external information (notated as $\lambda$ index) and $ \delta_i^{\lambda} = 0$ in other cases

\section{Hessian-Informed Regularization for Bayesian Peak Fitting}

We enhance prediction by applying the self-consistent Bayesian peak-fitting loop for DOS  by incorporating Hessian-based regularization derived from curvature analysis of the 1-layer GNN model. Specifically, we use the second derivative (Hessian) of the 1-layer prediction at critical points to define prior expectations for Gaussian peak amplitudes. The primary metric that will be utilized is the cosine distance and its Fourier transform. This is selected in such a manner as to maximize the distinction between predictions that identify and do not identify a peak at a particular point. This is fundamentally a direct consequence of the scalar product of the spectra.

The fundamental premise of the method is that the information obtained from the aforementioned analysis concerning potential peaks unrecognized in m-layer machine learning, as viewed through the lens of regularization theory, can be articulated as a priori information \ref{hessian}. This enables the incorporation of a term responsible for the peaks into the loss function. In this study, the focus will be constrained to the Gaussian representation.
The curvature prior imposes a soft constraint on amplitudes during optimization. The total loss function becomes:

\begin{align}\label{BHL}
\mathcal{L}_{\mathrm{BHL}} \;=\;
&\underbrace{ \left[ 1 - \cos\left( y_{\mathrm{enh}}, \hat{y}^{(3)} \right) \right] }_{\text{real‐space cosine}} 
&+ \lambda_f \underbrace{ \left[ 1 - \cos\left( F(y_{\mathrm{enh}}), F(\hat{y}^{(3)}) \right) \right] }_{\text{Fourier cosine}}
+ \lambda_A \underbrace{\mathbb{E} \left[ \left( \frac{A - A_{\mathrm{prior}}}{\sigma_A} \right)^2 \right] }_{\text{Hessian amplitude prior}} \notag\\
&+ \lambda_c \underbrace{\mathbb{E} \left[ \left| \Delta^2 y_{\mathrm{enh}} \right| \right] }_{\text{curvature penalty}}
\end{align}

where: $\hat{y}^{(3)}$ is the clear 3-layer GNN prediction.
	$y_{enh}$ is the enhanced prediction with Gaussian peaks added.
	$A_i^{\text{prior}} $ are amplitudes estimated from the curvature (i.e., negative second derivative) at critical points.
	$\sigma_{A_i}$  is a soft confidence width (hyperparameter).
	$\lambda $ controls the strength of the regularization.
The cosine distance between two spectral vectors $\mathbf{p}$ and $\mathbf{t}$ is defined as:

$$\mathcal{L}_{\text{cos}}(\mathbf{p}, \mathbf{t}) = 1 - \frac{\mathbf{p} \cdot \mathbf{t}}{||\mathbf{p}|| \cdot ||\mathbf{t}||}$$

Where $\mathbf{p} \cdot \mathbf{t}$ denotes the dot product, and $||\mathbf{p}||$ represents the L2 norm of vector $\mathbf{p}$.


$$\mathcal{L}_{\text{freq}}(\mathbf{p}, \mathbf{t}) = 1 - \frac{\mathcal{F}(\mathbf{p}) \cdot \mathcal{F}(\mathbf{t})}{||\mathcal{F}(\mathbf{p})|| \cdot ||\mathcal{F}(\mathbf{t})||}$$

Where $\mathcal{F}(\mathbf{p})$ denotes the magnitude of the Fourier transform of $\mathbf{p}$.


For analytical correctness in cases of weak smoothing of the original data, add a small technical term for curvature penalty $\Delta^2 y[n] = y[n+1] - 2y[n] + y[n-1]$.

\section{Results}\label{sec2}

To elucidate the matter at hand, it is imperative to commence with a delineation of the example Fig.2. The graph displays the following: the theoretical result (DFT), indicated by the black dotted line; the prediction line derived from the third layer of machine learning (green line); yellow dots representing the Gaussian peaks obtained through our post-processing procedure; and the line of the final prediction, indicating an enhancement (purple).

\begin{figure}[h!]\label{sample}
    \centering
    \includegraphics[width=0.95\linewidth]{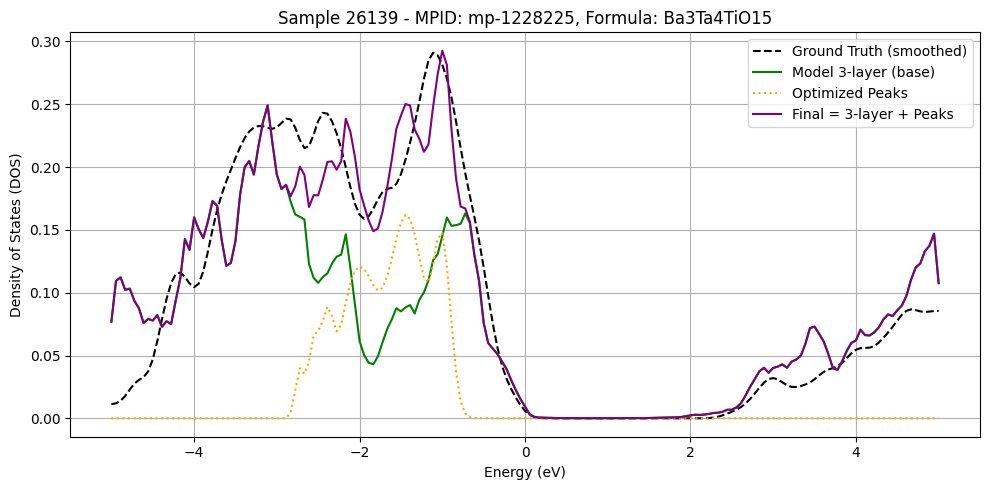}
    \caption{Sample of most effective recognition Ba3Ta4TiO15. Black line corresponds to smoothed Ground Truth; green line -  to clear 3-layer GNN prediction; orange points - to found Peaks; violet is final, normalized lineal combination of clear 3-layer GNN prediction and found peaks. }
    \label{fig:sample}
\end{figure}

This is a practical example of what improved prediction might look like under optimal conditions. In this case, the improvement occurred in almost all the metrics used. Unfortunately, the data is quite homogeneous in quality; moreover, the dataset used contains information about crystals, glasses, amorphous bodies, and even composites, the existence of which has not been proven.

 We shall proceed with the general scheme of application. The result will be most evident if the original 1-layer and m-layer models themselves were trained in the cosine metric. Here we apply only qVHS with positive amplitudes $A_i > 0$ (negative hessians in them). The results of this study are exclusively presented for the test dataset, with the use of smoothed model predictions ( $\sigma_{smooth} = pseudo = 2$) for post-processing. For the purpose of hyper-parameter optimization, the valid dataset was utilized exclusively. The efficacy of BHL approach is evidenced by its subsequent implementation Fig.3. The graph shows the dependence of the cosine distance metric on the number of training epochs for a 1-layer, 3-layer model and BHL.
One could see that the scale of improvement between the final results and the $m-$th layer results is comparable with the improvement between the $m-$th layer and the $1-$th layer ($m=3$).

.

\begin{figure}[h!]
    \centering
    \includegraphics[width=0.95\linewidth]{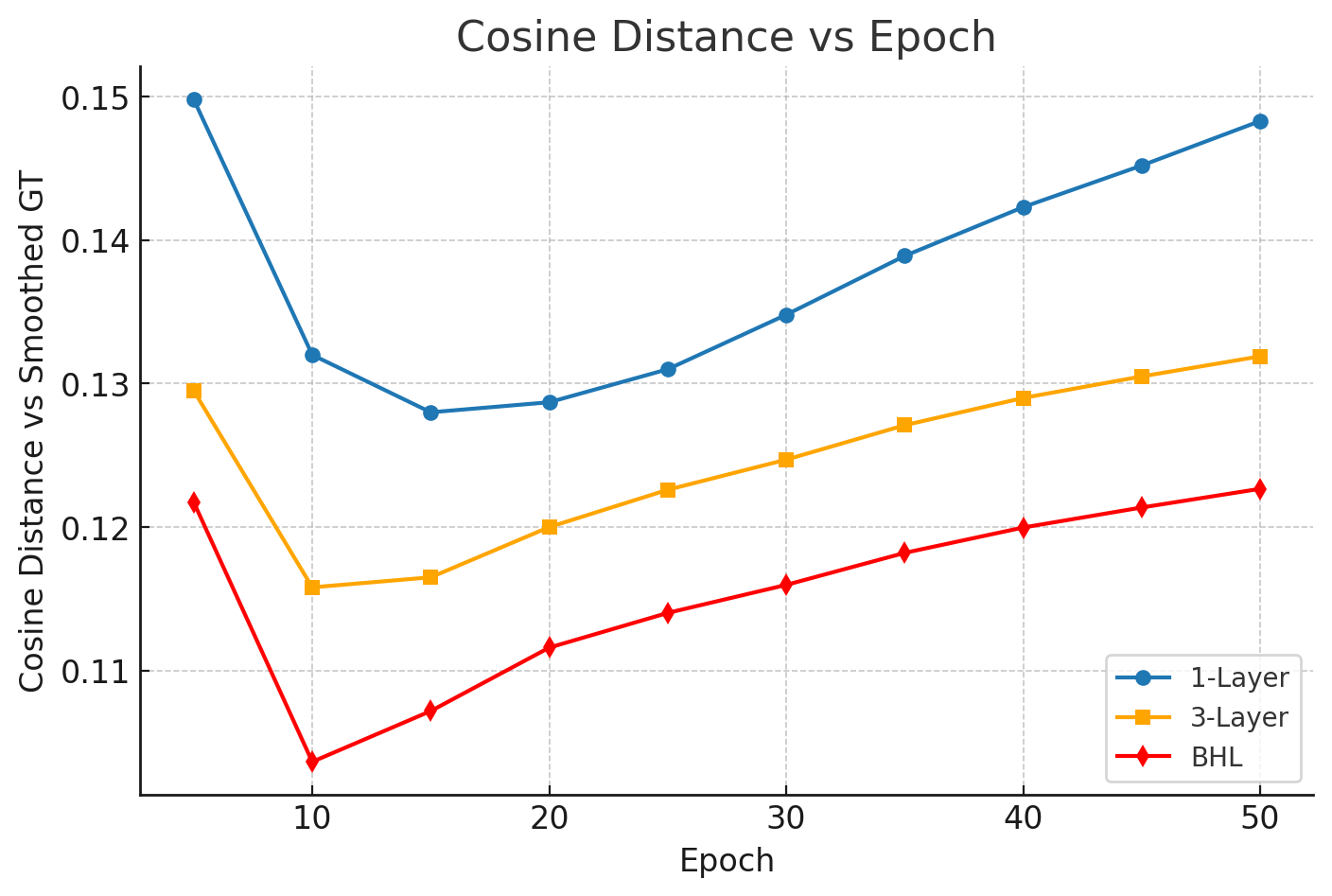}
    \caption{This plot compares the performance of various models and enhancement strategies for density of states prediction. The blue line represents a one-layer GNN model trained from scratch using the cosine distance metric. The orange line represents a three-layer GNN model trained from scratch using the cosine distance metric. The red line represents a post-processed correction using a typical Bayesian-Hessian enhancement.
}
    \label{fig:MAE}

\end{figure}

\subsection{NBHL normalization (normalozed BHL per atom)}

Since L1, L2 metrics are often highly favored in machine learning, we note that this method can also be applied to models fully trained in MAE. To achieve this objective, it is sufficient to incorporate a normalization operation into the absolute meaning of the $m-$th layer prediction integral within the specified energy range.

Let $y_{\mathrm{enh}}^{(a)}(E)$ denote the enhanced per-atom spectrum and $ y_{\mathrm{3}}^{(a)}(E) $ be the results of the 3-layer train.  
Define the normalization factor.
\begin{equation}
Z_a \;=\; \left| \int y_{\mathrm{3}}^{(a)}(E)\, dE \right|.
\end{equation}
Then the normalized enhanced spectrum is
\begin{equation}
\widehat{y}_{\mathrm{NBHL}}^{(a)}(E)
\;=\; \frac{y_{\mathrm{enh}}^{(a)}(E)}{Z_a},
\end{equation}
and the total normalized enhanced DOS (pDOS-to-DOS aggregation) reads
\begin{equation}
y_{\mathrm{NBHL}}^{\mathrm{tot}}(E)
\;=\; \frac{1}{{Z_a}}\sum_{a} y_{\mathrm{enh}}^{(a)}(E)
\qquad 
\end{equation}

In the event that NBHL is employed for models that have been trained in the MAE metric and subsequently presented for final comparison in the MAE, the relative effect of enhancing the prediction will be somewhat more modest than in the cosine-distance metric. However, it is statistically reliable.

\section{Conclusion}

In this paper, we put forth a novel approach that utilizes physical (or general numerical-physical priors) information to enhance the predictive capability of machine learning models. It has been demonstrated that a significant proportion of quantum systems exhibit manifestations of effects characteristic of the high-order VHS effect \cite{yuan2019magic}, \cite{chandrasekaran2024engineering}. It has been demonstrated that the implementation of these effects can be analytical in nature and tailored to the capabilities of machine learning. The logic underpinning this effect's prediction is predicated exclusively on the properties of differential mappings as applied to VHS-type systems. A potential subsequent action would be to undertake a more comprehensive examination of the correlation between the dispersion characteristics and machine learning methodologies. In this paper, we employed the predictions of the first layer as a source of clarifying information. However, the applied regularization method itself can also utilize other independent sources of information, such as measured spectra of various kinds or partial analytical calculations. In this instance, a refinement of the model will be necessary, which may facilitate a more precise prediction of Morse indices.


\section{Supplementary 1. Principal scheme of BHL approach}

\begin{lstlisting}[caption={Pseudo-Script for the Bayesian-Hessian Loss (BHL) Approach},label={lst:bhl_example}]
# Pseudo-Script: Bayesian-Hessian Loss (BHL) Approach for 
DOS Prediction Enhancement

# Inputs:
# - p1_all: 1-layer GNN DOS predictions, shape (N, L)
# - p3_all: 3-layer GNN DOS predictions, shape (N, L)
# - x_axis: Energy grid, [-5, 5] eV, shape (L,)
# - Hyperparameters: LAMBDA_REG, FREQ_WEIGHT, CURV_WEIGHT, INIT_SIGMA, 
MIN_SIGMA, MAX_SIGMA, PSEUDO_TARGET_SIGMA, FIT_LR, FIT_STEPS

# Output:
# - enh_pred_recal: Normalized enhanced DOS predictions, shape (N, L)

1. Initialize:
   - Set device (CUDA if available, else CPU)
   - Define energy grid: x_axis = linspace(-5, 5, L)
   - Load p1_all, p3_all (1-layer and 3-layer GNN predictions)
   - base = p3_all (baseline prediction)

2. Create Pseudo-Target:
   - pseudo = GaussianSmooth(base, sigma=PSEUDO_TARGET_SIGMA)
   # GT-free: pseudo is smoothed 3-layer prediction, not ground truth

3. Detect Quasi-Van Hove Singularities (qVHS):
   - For each sample i in p1_all:
      - spec1_i = p1_all[i] (1-layer DOS for sample i)
      - peaks_i = FindLocalMaxima(spec1_i) # Approximate zeros of dD/dE
      - mus_i = x_axis[peaks_i] # Peak positions
      - amps_i = spec1_i[peaks_i] # Initial amplitudes
      - sigmas_i = INIT_SIGMA * ones(size(peaks_i)) # Initial widths
   - Concatenate: musAll, ampsAll, sigmasAll, idx (sample indices)

4. Compute Hessian Priors:
   - For each sample i:
      - hess_i = GaussianSmooth(SecondDerivative(spec1_i), sigma=1.0)
      - prior_i = -hess_i[peaks_i] * INIT_SIGMA^2, clipped to >= 0
   - Concatenate: priorAll

5. Optimize Gaussian Peaks (BHL):
   - Initialize: ampsAll, sigmasAll as trainable parameters
   - Optimizer: Adam(ampsAll, sigmasAll, learning_rate=FIT_LR)
   - For step = 1 to FIT_STEPS:
      - recAll = GaussianSum(x_axis, idx, musAll, ampsAll, sigmasAll)
      - enh_pred = base + recAll
      - Loss = CosineDistance(enh_pred, pseudo)
            + FREQ_WEIGHT * FourierCosineDistance(enh_pred, pseudo)
            + LAMBDA_REG * Mean(((ampsAll - priorAll) / 0.5)^2)
            + CURV_WEIGHT * Mean(|SecondDifference(enh_pred)|)
      - Backpropagate Loss
      - Update ampsAll, sigmasAll
      - Constrain: ampsAll >= 0, MIN_SIGMA <= sigmasAll <= MAX_SIGMA

6. GT-Free Recalibration (Normalization):
   - I_pseudo = Integral(pseudo, x_axis) # Shape: (N,)
   - I_enh = Integral(enh_pred, x_axis) # Shape: (N,)
   - scale_factors = I_pseudo / (I_enh + 1e-8)
   - enh_pred_recal = enh_pred * scale_factors # Broadcast over energy axis

7. Evaluate:
   - Compute MAE_base = Mean(|p3_all - gt_smooth|)
   - Compute MAE_final_recal = Mean(|enh_pred_recal - gt_smooth|)
   - Compute MAE_base_pseudo = Mean(|p3_all - pseudo|)
   - Compute MAE_final_pseudo_recal = Mean(|enh_pred_recal - pseudo|)
   - Compute improvements:
      - Imp_MAE = 100 * (MAE_base - MAE_final_recal) / MAE_base
      - Imp_MAE_pseudo = 100 * (MAE_base_pseudo - MAE_final_pseudo_recal) /
      / MAE_base_pseudo

# Notes:
# - CosineDistance(p, t) = 1 - (p · t) / (||p|| · ||t||)
# - FourierCosineDistance(p, t) = CosineDistance(|FFT(p)|, |FFT(t)|)
# - GaussianSum computes sum of Gaussians for each sample
# - Integral uses trapezoidal rule
# - GT-free recalibration ensures physical consistency without ground truth
\end{lstlisting}




\end{document}